\documentclass{article}
\usepackage{amsfonts}
\usepackage{amssymb}
\usepackage{amsmath}
\usepackage{amsthm}
\usepackage{tensor}

\begin{document}

\title{\bf Ghosts in extended quasidilaton theories}

\author{Alexey Golovnev,  Aleksandr Trukhin\\
{\it Faculty of Physics, St. Petersburg State University,}\\ 
{\small\it Ulyanovskaya ul., d. 1, Saint Petersburg 198504, Russia.}
}

\date{}

\maketitle

\begin{abstract}

We report on our independent investigations of the puzzle of cosmological perturbations in extended quasidilaton. We confirm the claims of presence of the Boulware-Deser ghost. We use both the language of cosmological perturbations with broken diffeomorphisms and the St{\" u}ckelberg approach.

\end{abstract}

\section{Introduction}

The history of massive gravity dates back to the classical Fierz-Pauli model \cite{FP} in which it was shown that, out of two possible quadratic mass term for a metric fluctuation $h_{\mu\nu}\equiv g_{\mu\nu}-\eta_{\mu\nu}$ around  Minkowski spacetime, only one combination ($h_{\mu\nu}h^{\mu\nu}-(h_{\mu}^{\mu})^2$) gives healthy number of five degrees of freedom for a massive spin-two particle, whereas other options feature an extra mode with negative kinetic energy. Later it was realised that the problematic mode generically reappears after non-linear corrections have been taken into account, and it acquired the name of Boulware-Deser ghost after the work of Ref. \cite{BD}.

In recent years we have witnessed great progress in the theory of massive gravity. In particular, the healthy non-linear extension of the classical Fierz-Pauli model \cite{FP} has been found \cite{dRG,dRGT} and proven \cite{HR2,HR3,HR4} to be free of the Boulware-Deser ghost \cite{BD}. The theory is very peculiar in its mathematical features:  healthy potentials are elementary symmetric polynomials of eigenvalues of a square root of the matrix $g^{\alpha\mu}f_{\alpha\nu}$ where $f_{\mu\nu}$ is a fiducial metric which is often taken to be $\eta_{\mu\nu}$. Of course, beside the formal interest, there were natural hopes to explain the accelerated expansion of the Universe via Yukawa attenuation of the gravitational force at large distances. Unfortunately, not only didn't it work out properly, but even the very existence and stability of cosmological regimes was rather hard to achieve in massive gravity \cite{CosdRGT}.

It sparkled some interest towards extensions of massive gravity since what we have is a potentially healthy tensorial extension of general relativity which is very non-tirivial to achieve and can have interesting implications for cosmology. Popular options include bimetric gravity in which the fiducial metric is made dynamical, and also scalar-tensor extensions of two types: variable mass and quasidilaton \cite{quasidilaton}. The quasidilaton model features an additional dynamical scalar field $\sigma$, and the elemetary symmetric polynomials are calculated for the matrix $e^{\sigma/ M_\text{Pl}}\cdot\sqrt{g^{-1}f}$. Unfortunately, the cosmological perturbations turned out problematic \cite{qdpert}, and an extended version of this extension has been proposed \cite{extqd} which also shifts the fiducial metric by term proportional to $e^{-2 \sigma / M_\text{Pl}}
        \partial_\mu \sigma \partial_\nu \sigma$, see below.

The issue of stability in the full regime (as opposed to the so called late time limit) of extended quasidilaton presents a conundrum. In the first version of Shinji Mukohyama's preprint \cite{Shv1} the absence of the Boulware-Deser (BD) ghost has been proven with a particular gauge choice, which was however claimed to be not a good gauge \cite{Kluson}. In the work of Lavinia Heisenberg \cite{Lavinia} indications were given for stability of cosmological perturbations, at least in the far ultraviolet limit. Then, in the recent paper \cite{Glenn} it was claimed that cosmological perturbations have the BD ghost mode in the infrared limit contrary to previous believes. And the second version of the Mukohyama's preprint \cite{Shv2} proves that the BD ghost is present in the model.

We also had independent calculations of cosmological perturbations in quasidilaton which agree with the conclusions of Refs. \cite{Glenn,Shv2}. In Section 2 we briefly remind the formulation of extended quasidilaton model and its cosmological solutions. In Section 3 we present perturbative calculations in the language of cosmological variables {\it without} assuming infrared or ultraviolet limits. In Section 4 we arrive at the same results using the St{\" u}ckelberg fields. Finally, in Section 5 we discuss  and conclude.

\section{Cosmological solutions with extended quasidilaton}

Let us briefly set the stage for perturbative analysis of extended quasidilaton cosmologies. The action of extended quasidilaton model is given by
\begin{equation}
    S = \frac{M_\text{Pl}^2}{2} \int d^4x \sqrt{-g} \left( R[g] -
        \frac{\omega}{M_\text{Pl}^2} \partial_\mu \sigma \partial^\mu \sigma +
        2 m^2 \left( \mathcal{U}_2 + \alpha_3 \, \mathcal{U}_3 +
        \alpha_4 \, \mathcal{U}_4 \right)
    \right)
\end{equation}
where we have the usual massive gravity potential with elementary symmetric polynomials
\begin{subequations}
\begin{align}
    \mathcal{U}_2[K] &= \frac{1}{2} \, ([K]^2-[K^2]),\\
    \mathcal{U}_3[K] &= \frac{1}{6} \, ( [K]^3-3[K][K^2]+2[K^3] ) ,\\
    \mathcal{U}_4[K] &= \frac{1}{24} \, ( [K]^4 - 6[K]^2[K^2] + 3[K^2]^2 +
        8[K][K^3]-6[K^4] ) 
\end{align}
\end{subequations}
of the eigenvalues of the matrix
\begin{equation}
    K\indices{^\mu_\nu} = \delta\indices{^\mu_\nu} - e^{\sigma / M_\text{Pl}}
        \left( \sqrt{g^{-1} \tilde{f}} \right)\indices{^\mu_\nu}
\end{equation}
with the new fiducial metric given by
\begin{equation}
    \label{eq:a_sigma}
    \tilde{f}_{\mu\nu} = f_{\mu\nu} -
        \frac{a_\sigma}{M_\text{Pl}^2 m^2} e^{-2 \sigma / M_\text{Pl}}
        \partial_\mu \sigma \partial_\nu \sigma.
\end{equation}
Setting the quasidilaton field to $0$ gives the usual massive gravity. Vanishing of the new  coupling constant $a_\sigma$ while keeping the  $\sigma$ field arbitrary brings back the simple quasidilaton. Note that we neglect possible cosmological constant term in this action because it does not change our analysis, and anyway, one of the basic aims of those models is to obtain accelerated expansion {\it without} explicitly introducing a cosmological constant.

The fiducial metric is usually taken to be Minkowski: $f_{\mu\nu}=\eta_{\mu\nu}$. Note though that one can restore the diffeomorphism invariance by introducing the St{\" u}ckelberg fields $\phi^a$
\begin{equation}
    f_{\mu\nu} = \eta_{ab} \partial_\mu \phi^a \partial_\nu \phi^b
\end{equation}
which describe the change of coordinates from those in which the fiducial metric is Minkowski to arbitrary ones.

We are interested in the spatially flat FLRW solutions
\begin{align}
    \mathrm{d} s^2_g &= -N^2(t) \, dt^2 + a^2(t) \delta_{ij} \, dx^i dx^j,\\
    \mathrm{d} s^2_{\tilde{f}} &= -n^2(t) \, dt^2 + \delta_{ij} \, dx^i dx^j,
\end{align}
with $\sigma=\sigma(t)$. They correspond to
\begin{align}
    \phi_0 &= \phi_0(t),\quad \phi_i = x_i,\\
    n^2 &= \dot{\phi}_0^2 + \frac{a_\sigma}{M_\text{Pl}^2 m^2} e^{-2 \sigma / M_\text{Pl}}
        \dot{\sigma}^2. \label{eq:phi_0}
\end{align}
in the St{\" u}ckelberg language.

Background equations are easily found to be (see, e.g,~\cite{Lavinia})
\begin{equation}
    \label{eq:friedmann1}
    3H^2 = \Lambda_A + \frac{\omega\dot{\sigma}^2}{2M_\text{Pl}^2N^2},
\end{equation}
\begin{equation}
    \frac{2\dot{H}}{N} = \frac{(1-r)\dot{\Lambda}_A}{3HN-3\dot{\sigma}/M_\text{Pl}}
                         - \frac{\omega\dot{\sigma}^2}{M_\text{Pl}^2N^2},
\end{equation}
\begin{equation}
    \label{eq:stuckeq}
    \partial_t \left(
        \frac{1}{n} m^2 M_\text{Pl}^2 a^4 J(A-1)A  \, \dot{\phi}_0
    \right) = 0,
\end{equation}
\begin{equation}
    m^2 M_\text{Pl} N^3 A (3(r-1)A(-2+\alpha_3(A-1)) + J(-3+r(-1+4A))) 
    = \omega (3HN^2 \dot{\sigma} + N \ddot{\sigma} - \dot{N}\dot{\sigma})
\end{equation}
where we have introduced the following standard notations
\begin{subequations}
    \label{eq:qdhelper}
\begin{align}
    A(t) &\equiv \frac{e^{\sigma/M_\text{Pl}}}{a},\\
    H(t) &\equiv \frac{\dot{a}}{aN},\\
    r(t) &\equiv \frac{na}{N},\\
    J(t) &\equiv 3+3(1-A) \, \alpha_3+(1-A)^2 \, \alpha_4,\\
    \Lambda_A &\equiv m^2(A-1)[J+(A-1)(\alpha_3 \, (A-1)-3)].
\end{align}
\end{subequations}

\section{The ghost mode in cosmological perturbations}

Let us now study the cosmological perturbations. Since the BD ghost affects the scalar sector, we ignore vector and tensor perturbations. We use the following  parametrization of the scalar metric perturbations
\begin{subequations}
    \label{eq:gauge}
\begin{align}
    \delta g_{00} &= -2\, N^2\, \frac{\Phi}{M_\text{Pl}},\\
    \delta g_{0i} &= N \, a\, \partial_i \frac{B}{M_\text{Pl}},\\
    \delta g_{ij} &= a^2 \;\left[2 \delta_{ij} \frac{\psi}{M_\text{Pl}} + (\partial_i \partial_j -
    \frac{\delta_{ij}}{3} \partial^k \partial_k) \frac{E}{M_\text{Pl}}\right].
\end{align}
\end{subequations}

We consider quadratic expansion of the action around a cosmological solution.  Then we go to the Fourier space with spatial  momentum $k$, and get the quadratic action with the following relevant parts (for calculating the Hessian of the kinetic term):
\begin{eqnarray}
    \label{eq:l2}
    \mathcal{L}^{(2)} &=& -\frac{k^2 a^3 A B^2 Q}{2 (-1+A)^2 (1+r)}-2 k^2 a^2 B H \Phi
        -a^3 \Lambda_A \Phi ^2  \nonumber \\
    &&+\frac{1}{2} a^3 \left(\omega+\frac{a^2 (-1+A) J a_{\sigma } \dot{\phi }_0^2}{A r^3}\right) \delta\dot{\sigma
        }^2-\frac{1}{3} k^4 a^2 B \dot{E}+\frac{1}{12} k^4 a^3 \dot{E}^2 \nonumber \\
    &&-\omega a^3 \Phi  \delta \dot{\sigma } \dot{\sigma }+\left(2 k^2 a^2
        B+6 a^3 H \Phi \right) \dot{\psi } -3 a^3 \dot{\psi }^2  + \ldots
\end{eqnarray}
where
\begin{equation}
    \label{eq:qhelper}
    Q = -m^2 J (A-1) + (\Lambda_A + m^2 (A-1)^2)A.
\end{equation}
and, for the sake of simplicity, we have set the lapse to unity, $N=1$, and omitted the Planck mass $M_\text{Pl}$. Ellipsis at the end of Eq. (\ref{eq:l2}) shows that only those terms are written here which either are quadratic in velocities of dynamical variables or contain $\Phi$
or $B$ fields.

The fields $\Phi$
and $B$ can be eliminated by solving {\it algebraic} equations since their derivatives
do not enter the Lagrangian (we have included all  $\Phi$- and $B$-terms in~\eqref{eq:l2}).
After that we find
\begin{equation}
\begin{split}
    \label{eq:deth}
    \det{\mathcal{H}} \equiv
    \begin{vmatrix}
        \frac{\partial^2\mathcal{L}^{(2)}}{\partial\dot{\psi}^2} &
        \frac{\partial^2\mathcal{L}^{(2)}}{\partial\dot{\psi} \, \partial \dot{E}} &
        \frac{\partial^2\mathcal{L}^{(2)}}{\partial\dot{\psi} \, \partial(\delta\dot{\sigma})} &
        \\
        \frac{\partial^2\mathcal{L}^{(2)}}{\partial \dot{E} \, \partial\dot{\psi}} &
        \frac{\partial^2\mathcal{L}^{(2)}}{\partial \dot{E}^2} &
        \frac{\partial^2\mathcal{L}^{(2)}}{\partial \dot{E} \, \partial(\delta\dot{\sigma})} &
        \\
        \frac{\partial^2\mathcal{L}^{(2)}}{\partial(\delta\dot{\sigma}) \, \partial\dot{\psi}} &
        \frac{\partial^2\mathcal{L}^{(2)}}{\partial(\delta\dot{\sigma}) \, \partial \dot{E}} &
        \frac{\partial^2\mathcal{L}^{(2)}}{\partial(\delta\dot{\sigma})^2} &
    \end{vmatrix} = -\frac{
    k^4\, \omega\, a^{13} (-1+A)\, J \dot{\sigma }^2\, Q\, a_{\sigma } \dot{\phi }_0^2}
    {2 r^3 \left(2 k^2 (-1+A)^2 H^2 (1+r)-a^2 A Q \Lambda _A\right)}.
\end{split}
\end{equation}
It is easy to see that generically $\det{\mathcal{H}}\neq 0$ unless $a_{\sigma}=0$ (simple quasidilaton), or in the late time limit (the $J=0$ branch,~\cite{extqd}).
Therefore, there is the Boulware-Deser mode in the theory. 

It should also be noted that $\dot{\phi}_0 = 0$ is a singular case which does not correspond to the model with fixed fiducial metric. Indeed, the St{\" u}ckelberg fields describe the change of coordinates, and as such they must satisfy $\det \frac{\partial\phi^a}{\partial x^{\mu}}\neq 0$. Therefore, the $\dot{\phi}_0 = 0$ case from the Ref. \cite{Glenn} belongs to an unrestricted St{\" u}ckelberg model but not to the initial theory which has been covariantised by  St{\" u}ckelbergs.

Our ghost result is in accordance with the claims made in \cite{Glenn}. Moreover, their expressions can be easily obtained in the $k\to 0$ limit of our quadratic action. In most parts our formulae are also similar to those from Ref. \cite{Lavinia} which had the opposite conclusion. However, expressions from Ref. \cite{Lavinia} do lack terms with $a_{\sigma}$, presumably due to considerations of deep UV limit\footnote{Lavinia Heisenberg, private communication}, and setting this coupling to zero reduces the model to simple quasidilaton which is indeed BD-ghost-free. Note that taking a deep UV limit is a shaky ground for calculating the number of degrees of freedom since the Hessian might be non-degenerate at any finite wavenumber, but its different eigenvalues can have different $k\to\infty$ asymptotics which could lead to degeneracy in a simplified UV analysis.

\section{Treatment with St{\" u}ckelberg fields}

We have also checked the conclusion by calculations in the St{\" u}ckelberg picture. In this case we worked with slightly different formulation. In terms of $\beta$ coefficients instead of $\alpha$-s
the theory takes the form
\begin{equation}
    \label{eq:stuck_action}
    S = \frac{M_\text{Pl}^2}{2} \int d^4x \sqrt{-g} \left( R[g] -
        \frac{\omega}{M_\text{Pl}^2} \partial_\mu \sigma \partial^\mu \sigma +
        2 m^2 \sum_{n=0}^4 \beta_n \, \mathcal{U}_n[\mathbb{X}]
    \right)
\end{equation}
with
\begin{equation}
    \mathbb{X}\indices{^\mu_\nu} = e^{\sigma / M_\text{Pl}}
        \left( \sqrt{g^{-1} \tilde{f}} \right)\indices{^\mu_\nu}
\end{equation}
where $\mathcal{U}_0[\mathbb{X}] = 1$ and $\mathcal{U}_1[\mathbb{X}] = [\mathbb{X}]$.

This is an equivalent formulation since matrices $-\mathbb{X}$ and $K$ differ only by addition of the unit matrix, and therefore their elementary symmetric polynomials are related to each other. In particular, the $\beta_n$ coefficients should be taken as 
\begin{subequations}
    \label{eq:constredef}
\begin{align}
    \beta_0 &= 6 + 4 \alpha_3  + \alpha_4,\\
    \beta_1 &= -3 - 3 \alpha_3  - \alpha_4,\\
    \beta_2 &= 1 + 2 \alpha_3  + \alpha_4,\\
    \beta_3 &= -\alpha_3  - \alpha_4,\\
    \beta_4 &= \alpha_4
\end{align}
\end{subequations}
in terms of $\alpha_3, \alpha_4$ parameters from the previous section.

In the St{\" u}ckelberg picture we consider only perturbations of the quasidilton and of the St{\" u}ckelberg fields.
The relevant terms in the quadratic action  are those that
include $\dot{\delta\sigma}$, $\delta\dot{\phi}_a$ combinations (they do not kinetically mix with the physical metric)
\begin{eqnarray}
    \mathcal{L}^{(2)}  &=& \frac{e^{-3\sigma/M_\text{Pl}}}{2a^3m^2M_\text{Pl}^2n^3N^2(an+N)}\left[
    -2 e^{\frac{2 \sigma }{M_\text{Pl}}} m^2 M_\text{Pl} N (a n+N) \, \Theta \, \delta \dot{\sigma } \, \dot{\sigma } \, \delta\dot{
   \phi _0} \, \dot{\phi _0} \, a_{\sigma }\right.\nonumber\\
   &&\left.+(a n+N) \, \delta \dot{\sigma }^2 \left(a^3 e^{\frac{3 \sigma
   }{M_\text{Pl}}} m^2 M_\text{Pl}^2 n^3 \, \omega+e^{\frac{2 \sigma }{M_\text{Pl}}} m^2 M_\text{Pl}^2 n^2 N \, \Theta \, a_{\sigma }-N \, \Theta \,
   \dot{\sigma }^2 a_{\sigma }^2\right)\right.\nonumber\\
   &&\left.+e^{\frac{4 \sigma }{M_\text{Pl}}} m^4 M_\text{Pl}^2 N \left(
   (a n+N)\delta \dot{\phi _0}{}^2 \left(n^2-\dot{\phi _0}{}^2\right) \Theta
   - a n^3 \delta \dot{\phi _i}{}^2 \, \Xi \right)\right]
\end{eqnarray}
with
\begin{subequations}
\begin{align}
    \Theta &= a^3 \beta _1+3 a^2 e^{\frac{\sigma }{M_\text{Pl}}} \beta _2+3 a e^{\frac{2 \sigma }{M_\text{Pl}}} \beta _3+e^{\frac{3
   \sigma }{M_\text{Pl}}} \beta _4,\\
   \Xi &= a^3 \beta _1+2 a^2 e^{\frac{\sigma }{M_\text{Pl}}} \beta _2+ a e^{\frac{2 \sigma }{M_\text{Pl}}} \beta _3
\end{align}
\end{subequations}
where we have switched off the physical metric perturbations since now we are in the framework with restored diffeomorphism invariance, and the danger comes from the "matter" sector represented by St{\" u}ckelberg fields.

In the previous notations~\eqref{eq:qdhelper},~\eqref{eq:qhelper}
the final expression for $\mathcal{L}^{(2)}$ is
\begin{eqnarray}
    \mathcal{L}^{(2)} &=& \frac{1}{2 N^2} \left(
        \omega\, \delta\dot{\sigma }^2+ \frac{m^2 N}{a^3 A^3} \left( \frac{a^5 A^4 Q \delta \dot{\phi _i}{}^2}{m^2 (A-1)^2 (a n+N)}
    +\frac{a^4 (-1+A) A^4 \, J \, \delta \dot{\phi _0}{}^2
   \left(n^2-\dot{\phi _0}{}^2\right)}{n^3}\right.\right.\\
   &&\left.\left.-\frac{2 a^2 (-1+A) A^2 \, J \, \delta \dot{\sigma } \, \dot{\sigma } \,
   \delta \dot{\phi _0} \, \dot{\phi _0} \, a_{\sigma }}{m^2 M_\text{Pl} n^3}+\frac{(-1+A) J (\delta \dot{\sigma })^2
   a_{\sigma } \left(a^2 A^2 m^2 M_\text{Pl}^2 n^2-\dot{\sigma }^2 a_{\sigma }\right)}{m^4 M_\text{Pl}^2 n^3}
   \right) \right)   \nonumber
\end{eqnarray}

The corresponding Hessian is
\begin{align}
    \det \frac{\partial \mathcal{L}^{(2)}}{\partial\{\delta\dot{\sigma},
    \delta\dot{\phi}_a\}} = \frac{ \omega \, a_{\sigma} \, a^5 A^2 J Q^3 \dot{\sigma }^2}
            {(-1+A)^5 M_\text{Pl}^2 n^3 N^6 (a n+N)^3} \neq 0,
\end{align}
from which the presence of the B-D ghost is apparent.

\section{Discussion and Conclusions}

We confirm the presence of BD ghost in extended quasidilaton massive gravity which has been observed in Refs. \cite{Glenn,Shv2}.  A comment on the ghost-freedom "proof" of the paper \cite{Shv1} is in order. There the gauge choice of $\phi^0=-e^{-\sigma}$ has been used. It was noted in Refs. \cite{Kluson, Shv2} that it is not a correct gauge choice. We do not find a proper explanation there which is however quite simple. A gauge choice would amount to a condition which can be ensured by a coordinate transformation without restricting the physical variables. Obviously, with two arbitrary fields $\phi^0(t,x)$ and $\sigma(t,x)$ which can have different constant value surfaces in the spacetime, it is not possible to make one field a function of the other with simply a coordinate choice.

Now it seems firmly established that the model with extended quasidilaton is not ghost-free, and therefore it is not a viable option for massive cosmology. Though extremely compelling from the theoretical viewpoint, massive gravity is not doing as good for phenomenology (if not to play with bimetric regimes close to GR \cite{viable}). The quest for solving fundamental cosmological puzzles is as open as ever before.

{\bf Acknowledgements.} AG was supported by the Dynasty Foundation grant and
by Saint Petersburg State University research grant 11.38.223.2015. The authors are grateful to an anonymous referee who took the burden of checking the calculations and found some unpleasant mistakes in presentation of the main formulae.

\end{document}